\documentclass{INTERSPEECH2023}

\interspeechcameraready

\usepackage{amsmath}
\usepackage{graphicx}
\usepackage{booktabs}
\usepackage{tabularx}
\usepackage{url}
\usepackage{multirow,bigdelim}
\usepackage{pbox}
\usepackage[inkscapelatex=false]{svg}
\usepackage{hyperref}
\usepackage{enumitem}
\usepackage{microtype}

\hypersetup{
    colorlinks=true,
    citecolor=blue,
    linkcolor=blue,
    filecolor=magenta,      
    urlcolor=blue,
}

\newcommand{\ca}{\raisebox{0.5ex}{\texttildelow}}

\title{PLCMOS -- a data-driven non-intrusive metric for the evaluation of packet loss concealment algorithms}
\name{Lorenz Diener, Marju Purin, Sten Sootla, Ando Saabas, Robert Aichner, Ross Cutler}
\address{Microsoft Corporation}
\email{lorenzdiener@microsoft.com}

\begin{document}
%
\maketitle
\begin{abstract}
Speech quality assessment is a problem for every researcher working on models that produce or process speech. Human subjective ratings, the gold standard in speech quality assessment, are expensive and time-consuming to acquire in a quantity that is sufficient to get reliable data, while automated objective metrics show a low correlation with gold standard ratings. 

This paper presents PLCMOS, a non-intrusive data-driven  tool for generating a robust, accurate estimate of the mean opinion score a human rater would assign an audio file that has been processed by being transmitted over a degraded packet-switched network with missing packets being healed by a packet loss concealment algorithm. Our new model shows a model-wise Pearson's correlation of \ca0.97 and rank correlation of \ca0.95 with human ratings, substantially above all other available intrusive and non-intrusive metrics. 

The model is released as an ONNX model for other researchers to use when building PLC systems.
\end{abstract}

\section{Introduction}
\label{sec:intro}
Audio \emph{Packet Loss Concealment} (PLC) is the hiding of gaps in audio streams caused by data transmission failures in packet-switched networks. Every \emph{real-time communication} (RTC) system needs to solve this problem in some way -- the situation that there is no audio data available because a packet is lost or late is unavoidable in reality, and the system must still produce output. As voice-over-IP has increasingly become the default means of facilitating calls, PLC has consequently become an increasingly important concern~\cite{sun_impact_2001}.

With deep learning speech research making huge advances and edge compute becoming more powerful, it is natural to ask whether data-driven, neural techniques might be employed to improve upon classical PLC algorithms employed as part of RTC speech codecs. This is the field of Deep PLC. A big problem facing the Deep PLC research community is the problem of evaluation -- given two models, how can we tell which is better?

The obvious solution is to pass some data through the models and then have this data labeled by human listeners, either testing for preference directly, or having the listeners assign a \emph{mean opinion score} (MOS) rating. While, given enough raters, this approach will yield good results, it is unfortunately also expensive and slow and does not lend itself to quick iteration on new ideas that forms the basis of much research work. Consequently, it is generally only applied once much has already been settled, if at all. Other, automatically computable hand crafted metrics, such as PESQ~\cite{rix_perceptual_2001}, STOI~\cite{taal_algorithm_2011}, or MCD~\cite{kubichek_mel-cepstral_1993} are employed instead. These metrics, while quick and easy to compute, do not have a very strong correlation with human ratings, and may be insufficiently exact when trying to compare two relatively similar models. They also require an aligned reference, which limits their use to scenarios where such a reference is available. Notably, in case of packet loss concealment with a jitter buffer and timescale modification, which is the typical implementation, even if a reference exists, it is unaligned, and alignment may introduce additional errors. While other non-intrusive neural metrics~\cite{avila_non-intrusive_2019}, such as DNSMOS~\cite{reddy_dnsmos_2021}, AECMOS~\cite{purin2022aecmos} and NISQA~\cite{mittag21_interspeech} exist, they are more general or trained for other tasks, and may therefore not perform well on disambiguating the performance of similar PLC models.

In this paper, we present PLCMOS, a neural network trained to estimate the ratings human raters would assign to an audio file. Unlike the previously published PLCMOSv0 metric~\cite{diener2022interspeech}, which required a reference, PLCMOS is fully non-intrusive, while also performing better on the PLC MOS task. We present the dataset and model structure used and an evaluation and comparison with other metrics. The model is released in the Open Neural Network Exchange\footnote{ONNX Runtime developers, ``Onnx runtime'', https://onnxruntime.
ai/, 2021} format for use by other researchers\footnote{
\ifinterspeechfinal
\url{https://aka.ms/plcmos}
\else
redacted for review
\fi
}.

\section{Dataset}
\label{sec:dataset}

\subsection{Audio Data}
The audio data used in training is based on two datasets: The \textbf{LibriSpeech}~\cite{panayotov_librispeech_2015} dataset, and a LibriVox \textbf{Podcasts}\footnote{Librivox Contributors, ``The LibriVox community podcast'', https://librivox.org/
category/librivox-community-podcast/} dataset. The former is a collection of read speech from audiobooks, while the latter consists mostly of conversational speech taken from recordings of the LibriVox Podcast, which are in the public domain. For the purpose of PLCMOS, both of these are considered \emph{clean} speech, and to generate degraded audio, we process them by simulating packet loss. Audio data is primarily in English, with only very little non-English speech and non-verbal human sounds, as this is our primary research focus at this time. While we have not tested the model for non-English input, experience from previous models such as DNSMOS suggests that the model should still perform well. All audio was sampled at \SI{16}{\kilo\hertz}.

The audio data is converted to spectrograms using the short-term Fourier transform with a \SI{32}{\milli\second} Hamming window, a \SI{16}{\milli\second} frame shift, and the logarithm of the power is applied to generate the input features for our model. For training, we also perform microaugmentations (trimming up to 10 samples and reducing the volume by up to \SI{3}dBFS, both at random) to increase the robustness of our model against changes that a human would not interpret as strongly impacting quality. Audio segments are cut at random from base audio files with lengths matching the packet loss traces.

\subsection{Packet loss traces}
Many treatments of packet loss concealment use packet loss simulated using the two-state Gilbert model of packet loss~\cite{hasslinger2008gilbert}, however, such models are only accurate to a point~\cite{yu2005accuracy} and not appropriate as the basis for designing a data set used for PLC evaluation. Privacy concerns prevent sampling the audio data of full calls, and recording play-acted calls with a broad distribution of network conditions is prohibitively expensive. We, therefore, turn to real network packet traces recorded from 
\ifinterspeechfinal
Microsoft Teams calls.
\else
calls made using a popular RTC communications software.
\fi
 The traces record packet metadata such as losses and transmission times (though not the real audio data) for audio streams during these calls. By combining a trace with an audio file, we can obtain degraded audio that sounds exactly like a call using the given network transmission characteristics with the given audio would sound. We filter these traces to obtain more useful data for evaluation (see Section~\ref{sec:datasetbreakdown}). All data collection was cleared with internal privacy review teams.

\subsection{Sampling of packet loss traces}
Many calls experience only very light packet loss. Sampling equally from real packet loss traces would, therefore, result mostly in very easy cases that are not interesting or useful for the evaluation of PLC algorithms. To really cover the spectrum of what we should expect a PLC model to handle, it is necessary to filter data such that high loss and high burst cases are appropriately covered. As the basis for the PLCMOS dataset, we created two sets of traces:

\begin{description}
  \item [Basic]
    This trace set focuses on providing good coverage of realistic packet loss conditions.
    The traces were sampled as follows: First, 10-second segments with at least one lost packet were randomly extracted from the base trace set. From these, all segments with a burst loss of more than 120 milliseconds were discarded. The remaining traces were divided into 14 buckets according to packet percentage loss quantiles. Finally, an equal number of traces were sampled from each bucket, for a total of 1400 traces (100 per bucket). This gives a data set that we expect current Deep PLC models, which largely focus on short-term prediction, to perform well on.
        
  \item [Heavy loss]
    This trace set focuses on heavier and longer loss conditions, including losses we would generally consider to be irrecoverable by any real-time model.
    First, data were segmented as before. These segments were then divided into three subsets according to maximum burst loss length:

    \begin{itemize}
        \item Up to \SI{120}{\milli\second}
        \item Between \SI{120}{\milli\second} and \SI{320}{\milli\second}
        \item Between \SI{320}{\milli\second} and \SI{1000}{\milli\second}
    \end{itemize}

    Segments with burst losses longer than \SI{1000}{\milli\second} were discarded, as filling multi-second gaps at all is beyond the capability we currently expect of Deep PLC models (and even filling gaps up to a second is aspirational). Each subset was divided the same way as in the basic stratified set.  Finally, an equal number of traces were sampled from each bucket (with more traces being sampled per bucket for the subsets with shorter maximum burst losses).
        
  \item [Long bursts]        
    While the other sets focus on coverage of different rates of packet loss that may be more or less bursty, this trace set focuses specifically on long, but realistic burst losses, with loss rate being secondary. It was obtained by randomly sampling trace segments (obtained as in the previous two cases) that meet the following conditions:

    \begin{itemize}
        \item Maximum burst length between \SI{120}{\milli\second} and \SI{300}{\milli\second}
        \item Median burst length of at least \SI{80}{\milli\second}
        \item Packet loss percentage between 10 and 70 percent
    \end{itemize}

    A total of 500 traces were sampled for this dataset.
\end{description}

\subsection{PLC models}
To train a PLC MOS model, we require data that has been degraded through lossy transmission and then healed using PLC. By combining the audio data and the sampled traces, we can easily create lossy data, which we can then pass through either classical or neural PLC algorithms to create data that we can then have labeled by human raters to serve as a ground truth and training target. We used the following PLC methods to generate data to train and evaluate our PLCMOS model:

\begin{enumerate} \setlength\itemsep{-0.1em}
    \item No-op PLC / ``lossy" files (zero fill), oracle PLC
    \item Skype Silk and Satin codec PLC
    \item Google Lyra codec PLC
    \item Neural PLC model variants trained by Microsoft, employing different architectures (convolutional + recurrent, fully convolutional, E2E recurrent~\cite{thakker22_interspeech})
    \item Neural PLC models trained by participants of the INTERSPEECH 2022 Deep PLC Challenge~\cite{diener2022interspeech}
\end{enumerate}

Where possible, we employed these algorithms both with and without the use of additional packet loss concealment techniques such as jitter buffering and forward error correction, to create a dataset that is as diverse as possible.

\subsection{Labeling}

The ground truth scores we use for training and evaluating PLCMOS were obtained by using a crowd-sourcing approach based on the ITU P.808 framework~\cite{naderi_open_2020}. 

Audio clips were rated using the \emph{Absolute Category Rating} (ACR)~\cite{noauthor_p808_2021} approach. Raters were asked to assign discrete ratings ranging from 1 (Bad) to 5 (Excellent) for each degraded file and were not provided with a reference. As the PLC problem deals with dropouts in the speech signal, and we might expect people to be unable to tell speech dropped naturally apart from speech dropped due to packet loss, we also evaluated using a Comparison Category Rating approach, where both a degraded signal and a hidden reference are provided. However, after evaluating both approaches, we settled on ACR -- see the evaluation section for further consideration of this.

Raters were asked to evaluate the overall quality of a file and were asked to consider primarily how well they were able to understand the speaker. Raters were instructed to perform ratings in a quiet environment and to use headphones rather than loudspeakers. They were given some examples of files we would consider excellent (Score 5) and some files that we would consider bad (Score 1) to anchor raters expectations.

\subsection{Dataset Breakdown}

\label{sec:datasetbreakdown}

Combining audio datasets, model sets, and sampled traces in various ways, we obtain the dataset for training PLCMOS. Several models and a fraction of processed audio files were selected for evaluation and held out from training to evaluate the performance of the PLCMOS model on unseen PLC methods and audio. We add training data from the DNSMOS training set to increase robustness to general degradations beyond those caused by PLC. Table~\ref{tab:dataset} shows a breakdown of the dataset.

\begin{table}[ht]
\setlength{\tabcolsep}{5pt}
\centering
\caption{Composition of the PLCMOS dataset.}
\begin{tabular}{@{}llllll@{}}
\toprule
            &             & \multicolumn{2}{c}{\#Models} & \multicolumn{2}{c}{\#Votes} \\ \cmidrule(l){3-4} \cmidrule(l){5-6}
Audio data  & Trace set   & Train         & Eval         & Train         & Eval        \\ \midrule
LibriSpeech & Basic       & 78            & 21           & 333740        & 22165       \\
LibriSpeech & Long bursts & 10            & 2            & 15550         & 990         \\
Podcasts    & Heavy loss  & 17            &              & 82110         &             \\
DNSMOS      &             &               &              & 16800         &             \\ \bottomrule
\end{tabular}

\label{tab:dataset}

\end{table}

\section{Model Structure}

\label{sec:model}

The PLCMOS model consists of multiple modules. The input module takes audio input (in the form of log-pow spectrograms) and transforms them using a 3-layer convolutional encoder with 3x3 2D convolutions (with channel counts of 32, 64, and 64 and 2x dilation in the time direction) and a 4x reduction of the frequency dimension using max pooling. The resulting sequence is projected to 512 dimensions using a kernel width a 1D convolution, then passed through a bidirectional GRU layer. The final output of the embedding module is the final hidden states of the GRU layer. 

The ID embedding module takes an ID (See Sections \ref{sec:train} and \ref{sec:noid}) as input, picks a vector from a 64-dimensional normal distribution (constant per ID), and transforms it using a fully connected network (layer sizes [128, 128, 128, 128, 64]). During inference, we pick 25 random embeddings from a normal distribution as a ``virtual rater'' input. 

\begin{figure}
    \centering
    \includesvg[width=\columnwidth]{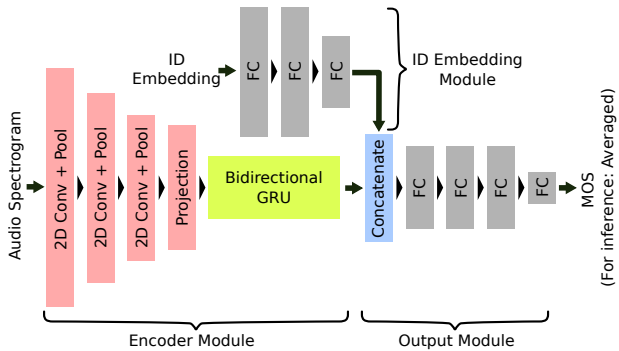}
    \caption{Structure of the PLCMOS neural network.}
    \label{fig:network}
\end{figure}

Module outputs are concatenated and then combined using a fully connected network (five layers, each layer containing 32 units except the final output layer, which contains just one). The output is transformed into the valid range for MOS ratings by linearly transforming the output of a sigmoid. Dropout with a rate of 0.1 is used between layers to avoid overfitting. Figure~\ref{fig:network} shows an overview of the network structure.

\subsection{Training} \label{sec:train}
The network was trained for 250 epochs using a batch size of 16 and a learning rate of 0.0003 with an AdamW optimizer (beta values 0.9 and 0.999) optimizing the mean squared error between predicted and actual rating. Various model configurations (No dilation in convolutions or dilation size of 2, aggregation of values using simple mean or bidirectional GRU, number of layers in ID embedding network -- 3 or 5, inclusion of an additional self-attention block before aggregation) for the network were tested during model development. The final model configuration was selected according to best rank correlation on the evaluation holdout.

Ratings were not aggregated to MOS for training -- instead, we allow the network to predict a range of scores by providing it with a rater ID, provided as an additional normally distributed input, based on the assumption that raters make systematic errors, and that ratings for a file between raters are normally distributed. Because rater IDs aren't available for all votes in the dataset -- specifically, they are unavailable for DNSMOS data -- PLCMOS uses a vote ID as input in those cases (i.e. assumes distinct raters for every vote). During evaluation, when a rater ID is of course not available, we sample a set of virtual raters from a standard normal distribution and average the scores obtained this way.

In addition to the evaluation model using the holdout described in Section~\ref{sec:datasetbreakdown} we also train and compare an additional model with the ID input omitted for evaluation (See Section~\ref{sec:noid}) and a no-holdout release model.

\section{Evaluation}
\subsection{ACR vs. CCR labeling for reference data}
Given that PLC deals with the repairing of dropouts, using a method such as the P.808 \emph{Comparison Category Rating} (CCR)~\cite{noauthor_p808_2021} where a reference is available seems potentially beneficial -- after all, without a reference, a rater might not be able to tell that part of an audio file is missing, especially in a crowd-sourcing setting with amateur raters. To test this, we performed a pre-study: As ground truth, we had the same set of 620 lossy clips CCR-labeled by members of our team listening for PLC-related issues, and by crowd-sourcing raters using CCR and ACR.

We compare how well crowdsource ratings for the files correlate with the expert CCR ratings, and find a \ca0.83 (95\% CI: [0.78, 0.87]) PCC for crowd-sourced ACR, and \ca0.86 (95\% CI: [0.83, 0.89]) for crowd-sourced CCR. This difference is not significant. Given the practical advantages of ACR over CCR (ability to have reference-free files rated, higher throughput / lower cost, as raters only have to listen to one clip for each rating), we used an ACR approach for our ground truth labeling.

\label{sec:eval}
\subsection{PLCMOS vs. other metrics}

\begin{table}[t]
\setlength{\tabcolsep}{3pt}
\centering
\caption{PLCMOS vs. other neural metrics, full evaluation set}
\begin{tabular}{@{}lllllll@{}}
\toprule
                & \multicolumn{3}{c}{Filewise} & \multicolumn{3}{c}{Modelwise} \\ \cmidrule(l){2-4} \cmidrule(l){5-7} 
Metric          & PCC           & SRCC  & MAE        & PCC     & SRCC & MAE         \\ \midrule
DNSMOS          & 0.52          & 0.45  & 0.71       & 0.85    & 0.68 & 0.37        \\
NISQA (MOS)     & 0.69          & 0.66  & 0.67       & 0.81    & 0.71 & 0.47        \\
NISQA (DIS)     & 0.63          & 0.63  & 0.72       & 0.66    & 0.66 & 0.51        \\
PLCMOSv0        & 0.81          & 0.79  & 0.48       & 0.94    & 0.92 & 0.29        \\
PLCMOS (no ID)  & 0.83          & 0.80  & 0.45       & 0.95    & 0.95 & 0.20        \\
PLCMOS (ours)   & \textbf{0.85} & \textbf{0.83} & \textbf{0.40} & \textbf{0.97} & \textbf{0.95} & \textbf{0.09}        \\ \bottomrule
\end{tabular}
\label{tab:neural}
\end{table}

We evaluate the metrics according to how well ratings from the metric correlate with the ground truth MOS ratings. We evaluate the \emph{Pearson correlation coefficient} (PCC), \emph{Spearman rank correlation coefficient} (SRCC) and \emph{mean absolute error} (MAE, not normalized, score range 1 - 5) for both single files (averaging over votes for that file, labeled \emph{filewise}) as well as averaged over all files for a given model (labeled \emph{modelwise}). We consider the modelwise SRCC the most important, as it indicates how well a metric can be used to rank PLC algorithms according to performance. None of the trace patterns, audio, or PLC models used for testing were part of the PLCMOS training set.

\begin{figure}
    \centering
    \includesvg[width=\columnwidth]{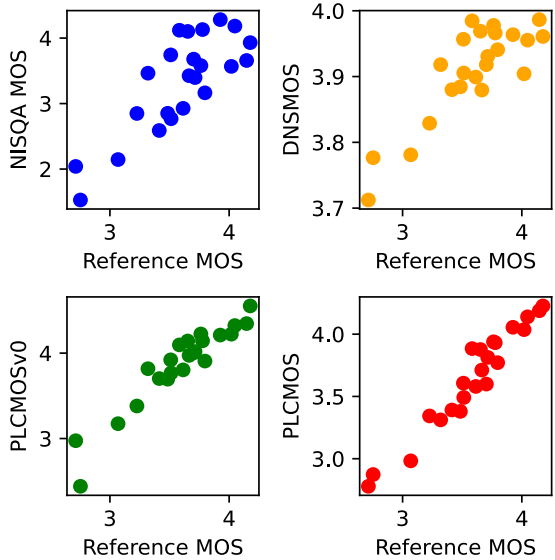}
    \caption{Scatterplots illustrating correlation of neural metrics and reference score. Axes scaled to cover range of data.}
    \label{fig:scatter}
\end{figure}

We compare our PLCMOS with two other types of metrics. First, with other non-intrusive metrics (DNSMOS, NISQA) and PLCMOSv0~\cite{diener2022interspeech}, a model that uses a non-aligned reference. For NISQA, we compare both overall MOS and Discontinuity score. We also compare a version of PLCMOS that does not use an ID input (see Section~\ref{sec:noid}). Table~\ref{tab:neural} shows these results. Our model beats both the more general NISQA model as well as DNSMOS and PLCMOSv0 on the PLC MOS task by a large margin. We better correlations with the ground truth for the overall NISQA score than the Discontinuity score, and similar correlations for DNSMOS with. 
Figure~\ref{fig:scatter} illustrates the improved correlation with scatter plots of modelwise metric vs. reference scores for the neural models we evaluated.

Next, we compare our model to more classical metrics. For this, we only use the part of the test set where an aligned reference is available (i.e., we exclude all data where timescale modification was used as part of the PLC process). The results of this investigation can be seen in Table~\ref{tab:classic}. We omit the MAE, as ranges of these metrics differ. The classical metrics only partially capture the degradations in quality -- even the best metric in our evaluation only achieves a modelwise SRCC of \ca0.54 compared to \ca0.97 for PLCMOS

\subsection{Ablation: Impact of ID embedding} \label{sec:noid}
We test how well our model functions when training without the ID embedding input and predicting the MOS directly instead of multiple votes (``PLCMOS (no ID)" in Table~\ref{tab:neural}). We find that while modelwise SRCC stays about the same, all other metrics degrade, indicating greater stability for the model that can make use of the IDs in training.


\subsection{Potential limitations}
There are several potential limitations of our PLCMOS model as released that users should be aware of.

\begin{description}
  \item [Sample Rate] While there has been a trend towards going beyond wide-band audio in real time communications, the PLCMOS model has been trained only on \SI{16}{\kilo\hertz} data. While evaluation by down-sampling before passing data to the model is possible, this usage has not been validated. The same goes for upsampling lower bandwidth audio data.
  \item [Language and mode] PLCMOS is trained, for the most part, on English modal speech. While we expect similar degradations for other languages, other modes of speech (e.g.~whispered speech, impaired speech) or even non-speech audio, this usage is out of domain for the current PLCMOS model and has not been validated.
\end{description}

\subsection{Final release model}
For practical use, to maximize coverage, we train a final release model that is structurally identical to the evaluation model presented in this paper, but which has been trained on the entire dataset with no holdout. We do not show metrics for this model, as it is trained with no holdout and such metrics would therefore be meaningless. 

The model contains a total of 299265 parameters and was trained on an Azure ND40rsv2 compute node with 8 NVIDIA Tesla V100 GPUs for \ca\SI{105}{\hour} (250 epochs, \ca353516k samples), using \ca\SI{114}{\kilo\watt\hour} of energy for compute (as estimated by AzureML).

\begin{table}[t]
\centering
\caption{Comparison of PLCMOS with classical metrics, dataset restricted to only data with aligned reference.}
\begin{tabular}{@{}lllll@{}}
\toprule
                    & \multicolumn{2}{c}{Filewise} & \multicolumn{2}{c}{Modelwise} \\ \cmidrule(l){2-3} \cmidrule(l){4-5} 
Metric              & PCC           & SRCC         & PCC           & SRCC          \\ \midrule
MCD           &  0.14 &  0.21 &  0.23 &  0.06 \\ 
PESQ          &  0.70 &  0.76 &  0.52 &  0.54 \\
STOI          &  0.03 &  0.17 &  0.21 &  0.26 \\ 
PLCMOS (ours) &   \textbf{0.87} &  \textbf{0.85} &  \textbf{0.98} &  \textbf{0.97} \\
\bottomrule
\end{tabular}
\label{tab:classic}
\end{table}

\section{Conclusion}
\label{ssec:conclusion}
We have presented PLCMOS, a neural model for evaluating the quality of PLC algorithm output audio. Unlike previous metrics used for this, it is non-intrusive, greatly expanding its practical usability. We have shown that the model compares favorably to these metrics, achieving a \ca.95 SRCC on the evaluation set, meaning that our new metric is substantially better at ranking PLC algorithms than other metrics. We have released this model (
\ifinterspeechfinal
\url{https://aka.ms/plcmos}
\else
redacted for review
\fi
) to allow other researchers to use it and help accelerate Deep PLC research.

In the future, we plan to investigate how to better exploit information about raters, to add more and more diverse ratings to our data set to further improve model performance and reliability, and to validate this performance on more models trained to work in a wider variety of packet loss conditions as well es on higher bandwidth audio files.

\bibliographystyle{IEEEtran}
\bibliography{main}

\begin{thebibliography}{10}
\providecommand{\url}[1]{#1}
\csname url@samestyle\endcsname
\providecommand{\newblock}{\relax}
\providecommand{\bibinfo}[2]{#2}
\providecommand{\BIBentrySTDinterwordspacing}{\spaceskip=0pt\relax}
\providecommand{\BIBentryALTinterwordstretchfactor}{4}
\providecommand{\BIBentryALTinterwordspacing}{\spaceskip=\fontdimen2\font plus
\BIBentryALTinterwordstretchfactor\fontdimen3\font minus
  \fontdimen4\font\relax}
\providecommand{\BIBforeignlanguage}[2]{{%
\expandafter\ifx\csname l@#1\endcsname\relax
\typeout{** WARNING: IEEEtran.bst: No hyphenation pattern has been}%
\typeout{** loaded for the language `#1'. Using the pattern for}%
\typeout{** the default language instead.}%
\else
\language=\csname l@#1\endcsname
\fi
#2}}
\providecommand{\BIBdecl}{\relax}
\BIBdecl

\bibitem{sun_impact_2001}
L.~F. Sun, G.~Wade, B.~M. Lines, and E.~C. Ifeachor, ``Impact of {Packet}
  {Loss} {Location} on {Perceived} {Speech} {Quality},'' in \emph{In 2nd
  {IP}-{Telephony} {Workshop}}, 2001, pp. 114--122.

\bibitem{rix_perceptual_2001}
\BIBentryALTinterwordspacing
A.~Rix, J.~Beerends, M.~Hollier, and A.~Hekstra, ``Perceptual evaluation of
  speech quality ({PESQ})-a new method for speech quality assessment of
  telephone networks and codecs,'' in \emph{2001 {IEEE} {International}
  {Conference} on {Acoustics}, {Speech}, and {Signal} {Processing} ({ICASSP
  2001})}, vol.~2.\hskip 1em plus 0.5em minus 0.4em\relax Salt Lake City, UT,
  USA: IEEE, 2001, pp. 749--752. [Online]. Available:
  \url{http://ieeexplore.ieee.org/document/941023/}
\BIBentrySTDinterwordspacing

\bibitem{taal_algorithm_2011}
C.~H. Taal, R.~C. Hendriks, R.~Heusdens, and J.~Jensen, ``An {Algorithm} for
  {Intelligibility} {Prediction} of {Time}–{Frequency} {Weighted} {Noisy}
  {Speech},'' \emph{IEEE Transactions on Audio, Speech, and Language
  Processing}, vol.~19, no.~7, pp. 2125--2136, Sep. 2011.

\bibitem{kubichek_mel-cepstral_1993}
R.~Kubichek, ``Mel-cepstral distance measure for objective speech quality
  assessment,'' in \emph{Proceedings of {IEEE} {Pacific} {Rim} {Conference} on
  {Communications} {Computers} and {Signal} {Processing}}, vol.~1, May 1993,
  pp. 125--128 vol.1.

\bibitem{avila_non-intrusive_2019}
\BIBentryALTinterwordspacing
A.~R. Avila, H.~Gamper, C.~Reddy, R.~Cutler, I.~Tashev, and J.~Gehrke,
  ``\BIBforeignlanguage{en}{Non-intrusive speech quality assessment using
  neural networks},'' \emph{\BIBforeignlanguage{en}{2019 IEEE International
  Conference on Acoustics, Speech and Signal Processing (ICASSP 2019)}}, Mar.
  2019, arXiv: 1903.06908. [Online]. Available:
  \url{http://arxiv.org/abs/1903.06908}
\BIBentrySTDinterwordspacing

\bibitem{reddy_dnsmos_2021}
\BIBentryALTinterwordspacing
C.~K.~A. Reddy, V.~Gopal, and R.~Cutler, ``{DNSMOS}: {A} {Non}-{Intrusive}
  {Perceptual} {Objective} {Speech} {Quality} metric to evaluate {Noise}
  {Suppressors},'' \emph{2021 IEEE International Conference on Acoustics,
  Speech and Signal Processing (ICASSP 2021)}, Feb. 2021, arXiv: 2010.15258.
  [Online]. Available: \url{http://arxiv.org/abs/2010.15258}
\BIBentrySTDinterwordspacing

\bibitem{purin2022aecmos}
M.~Purin, S.~Sootla, M.~Sponza, A.~Saabas, and R.~Cutler, ``Aecmos: A speech
  quality assessment metric for echo impairment,'' in \emph{2022 IEEE
  International Conference on Acoustics, Speech and Signal Processing ({ICASSP
  2022})}.\hskip 1em plus 0.5em minus 0.4em\relax IEEE, 2022, pp. 901--905.

\bibitem{mittag21_interspeech}
G.~Mittag, B.~Naderi, A.~Chehadi, and S.~Möller, ``{NISQA: A Deep
  CNN-Self-Attention Model for Multidimensional Speech Quality Prediction with
  Crowdsourced Datasets},'' in \emph{22nd Annual Conference of the
  International Speech Communication Association ({INTERSPEECH 2021})}, 2021,
  pp. 2127--2131.

\bibitem{diener2022interspeech}
L.~Diener, S.~Sootla, S.~Branets, A.~Saabas, R.~Aichner, and R.~Cutler,
  ``Interspeech 2022 audio deep packet loss concealment challenge,'' in
  \emph{23rd Annual Conference of the International Speech Communication
  Association ({INTERSPEECH 2022})}, 2022.

\bibitem{panayotov_librispeech_2015}
\BIBentryALTinterwordspacing
V.~Panayotov, G.~Chen, D.~Povey, and S.~Khudanpur, ``Librispeech: {An} {ASR}
  corpus based on public domain audio books,'' in \emph{2015 {IEEE}
  {International} {Conference} on {Acoustics}, {Speech} and {Signal}
  {Processing} ({ICASSP 2015})}.\hskip 1em plus 0.5em minus 0.4em\relax South
  Brisbane, Queensland, Australia: IEEE, Apr. 2015, pp. 5206--5210. [Online].
  Available: \url{http://ieeexplore.ieee.org/document/7178964/}
\BIBentrySTDinterwordspacing

\bibitem{hasslinger2008gilbert}
G.~Ha{\ss}linger and O.~Hohlfeld, ``The gilbert-elliott model for packet loss
  in real time services on the internet,'' in \emph{14th GI/ITG
  Conference-Measurement, Modelling and Evalutation of Computer and
  Communication Systems}.\hskip 1em plus 0.5em minus 0.4em\relax VDE, 2008, pp.
  1--15.

\bibitem{yu2005accuracy}
X.~Yu, J.~W. Modestino, and X.~Tian, ``The accuracy of gilbert models in
  predicting packet-loss statistics for a single-multiplexer network model,''
  in \emph{Proceedings IEEE 24th Annual Joint Conference of the IEEE Computer
  and Communications Societies.}, vol.~4.\hskip 1em plus 0.5em minus
  0.4em\relax IEEE, 2005, pp. 2602--2612.

\bibitem{thakker22_interspeech}
M.~Thakker, S.~E. Eskimez, T.~Yoshioka, and H.~Wang, ``{Fast Real-time
  Personalized Speech Enhancement: End-to-End Enhancement Network (E3Net) and
  Knowledge Distillation},'' in \emph{23rd Annual Conference of the
  International Speech Communication Association ({INTERSPEECH 2022})}, 2022,
  pp. 991--995.

\bibitem{naderi_open_2020}
\BIBentryALTinterwordspacing
B.~Naderi and R.~Cutler, ``An {Open} {Source} {Implementation} of {ITU}-{T}
  {Recommendation} {P}.808 with {Validation},'' \emph{21st Annual Conference of
  the International Speech Communication Association ({INTERSPEECH 2020})}, pp.
  2862--2866, Oct. 2020, arXiv: 2005.08138. [Online]. Available:
  \url{http://arxiv.org/abs/2005.08138}
\BIBentrySTDinterwordspacing

\bibitem{noauthor_p808_2021}
{International Telecommunications Union}, ``Subjective evaluation of speech
  quality with a crowdsourcing approach,'' \emph{ITU-T Recommendation P.808},
  2021.

\end{thebibliography}

\end{document}